\documentstyle[12pt]{article}
\newcommand{\bea}{\begin{eqnarray}}
\newcommand{\eea}{\end{eqnarray}}
\newcommand{\be}{\begin{equation}}
\newcommand{\ee}{\end{equation}}
\newcommand{\vs}[1]{\vspace{#1 mm}}

\newcommand{\dsl}{\pa \kern-0.5em /}

\newcommand{\pa}{\partial}

\newcommand{\nn}{\nonumber\\}

\begin{document}
\topmargin 0pt
\oddsidemargin 0mm

\begin{flushright}
hep-th/0308069\\
\end{flushright}

\vs{2}
\begin{center}
{\Large \bf  
On the effective action of a space-like brane}
\vs{10}

{\large Somdatta Bhattacharya$^a$,
 Sudipta Mukherji$^b$ and Shibaji Roy$^a$}
\vspace{5mm}

{\em 
 $^a$ Saha Institute of Nuclear Physics,
 1/AF Bidhannagar, Calcutta-700 064, India\\
 E-Mails: som, roy@theory.saha.ernet.in\\

\vs{2}

$^b$ Institute of Physics, Bhubaneswar 751 005, India\\
 E-Mail: mukherji@iopb.res.in}
\end{center}

\vs{5}
\centerline{{\bf{Abstract}}}
\vs{5}
\begin{small}
Starting from the non-BPS D$(p+1)$-brane action, we derive an effective 
action in $(p+1)$ space dimensions by studying the fluctuations of various 
bosonic fields around the time-like tachyonic kink solution (obtained by 
Wick rotation of the space-like tachyonic kink solution) of the non-BPS 
brane. In real time this describes the dynamics of a space-like or 
Euclidean brane in $(p+1)$ dimensions containing a Dirac-Born-Infeld (DBI) 
part and an Wess-Zumino (WZ) part. The WZ part is purely imaginary and so 
the action is complex if it represents the source of the time-dependent 
background of type II string theory i.e. the S-brane. On the other hand, 
the  WZ part as well as the action  is real if it represents the source in 
type II$^\ast$ string theory i.e. the E-brane. The DBI part is the same as 
obtained before using different method. This is then further illustrated by 
considering brane probe in space-like brane background.
\end{small}
\newpage

Non-BPS D$(p+1)$-branes exist
in both type IIA (for $p$ = even) and type IIB (for $p$
= odd)
string theory and are unstable due to the presence
of open string tachyon in their world-volume \cite{asone}. The dynamics of 
the non-BPS
branes can be described by an effective tachyon field theory 
\cite{astwo,mg,bddep,jk} and if the
tachyon depends on one of the space-like coordinates ($x^{p+1} \equiv x$
(say)) of the brane, it has an infinitely thin but finite tension kink
solution \cite{asthree}. It has been shown that for this solution the 
total energy of the
brane is localized around $x=0$ with the tension ${\cal T}_p 
= \int V(T)dT$, $T$ being the tachyon. Sen
has shown \cite{asfour} that even though the tachyon effective action 
is valid (where 
higher ($\geq 2$) derivatives of tachyon are neglected) for large $T$, the
fluctuations of massless modes on the kink solution interpolating between
the two vacua at $T=-\infty$ and $T=\infty$, passing through $T=0$, correctly
reproduce the DBI action of the BPS D$p$-brane without any higher derivative 
corrections. Also, the fluctuations are not assumed to be small in this
derivation. The minimum of the potential $V(T)$ has been argued to describe
the closed string vacuum \cite{asone}.

In order to understand the decay of the non-BPS D$(p+1)$-brane, one should
really
consider the tachyon to be time dependent \cite{asfive}. Thus the rolling 
of the tachyon
is responsible for the decay of the non-BPS D$(p+1)$-brane to the closed string
vacuum \cite{assix}. On the other hand, by a similar reasoning as 
given in the previous
paragraph, the rolling of the tachyon can also be seen to be responsible
for the appearance of the space-like $p$-branes \cite{gs} (S$p$-branes or 
E$p$-branes)
at the maximum
of the tachyon potential $V(T)$ where $T \to 0$. S$p$-branes (E$p$-branes)
are space-like
topological defects localized in $(p+1)$-dimensional space-like hypersurface
in the type II (type II$^\ast$)\footnote{Type II$^\ast$ string theory 
is related 
to type II string theory by a time-like T-duality \cite{ch}. So, for example, 
type IIA
(type IIB) string theory compactified on a time-like circle of radius $R$ is
dual to type IIB$^\ast$ (type IIA$^\ast$) string theory compactified on a dual
time-like circle of radius $1/R$. Space-like $p$-branes in type II theory
are the S$p$-branes and the space-like branes in type II$^\ast$ theory are
the E$p$ branes.}
string theory
and appears when finely tuned incoming closed string radiation pushes the
tachyon (at $x^0 = -\infty$) to the top of the potential from one side and 
then the tachyon rolls down to the other side (at $x^0 = \infty$) dissipating
its energy back to radiation \cite{msy}. In this process, the space-like 
$p$-brane appears and exists only for a moment in time at $x^0=0$. These 
are the sources for the classical
time dependent solutions \cite{td,cgg,kmp,sr} of the string effective 
action\footnote{Actually here we
are considering space-like $p$-brane solutions without a time reversal 
symmetry. It is also possible to consider solutions having a time 
reversal symmetry. A world-sheet construction for a particular 
boundary interaction
(corresponding to $\lambda=1/2$) \cite{asfive} describing the tachyon 
on an unstable D-brane
reveals that there is an array of space-like $p$-branes 
situated on the imaginary time axis at $x^0 = im\pi$, for odd integers $m$. 
In real time this has been interpreted as closed string radiation 
\cite{msy,llm,gir}.}.

In this paper, we give a simple derivation of a S$p$-brane (or E$p$-brane)
action starting 
from the non-BPS D$(p+1)$-brane action with time dependent tachyon. Here we
make use of Sen's derivation \cite{asfour} of a codimension one BPS D-brane 
action as the
space-like tachyonic kink solution of the non-BPS D-brane action. When the
tachyon is time dependent, the solution of the equation of motion can be
obtained by a Wick rotation of the corresponding solution with space-dependent
tachyon. This solution may be regarded as the time-like tachyonic kink of the
non-BPS D-brane action. We then study the fluctuations of various bosonic
fields on the Euclidean world-volume around this solution and obtain an 
effective action. In real time, this effective action represents the 
S$p$-brane (or E$p$-brane) action describing the dynamics of the various 
fields living on
the brane. We derive both the DBI and the WZ parts of the action, where the
DBI part matches with the results obtained earlier \cite{hhw} using 
different method. We
will point out in what sense we call this an S$p$-brane or an E$p$-brane 
action. We also give an alternative argument to further support the form
of the action that we have obtained. This is done
 by considering the brane probe in a non-extremal $p$-brane
background. The $p$-branes turn into  space-like branes
as we consider the region beyond their horizons. The probe action,
continued beyond the horizon, is found to take the form of the space-like
brane action that we have been discussing so far.
 
The dynamics of the non-BPS D$(p+1)$-brane is governed by the following 
tachyon effective action\footnote{We are using the convention where
$\eta_{\mu\nu} = {\rm diag}(-1,1,\ldots,1)$ and $\alpha'=1$.}
\cite{astwo,mg,bddep,jk},
\be
S = - \int d^{p+2}x V(T) \sqrt{-{\rm det}(\eta_{\mu\nu} + \partial_\mu T
\partial_\nu T)}
\ee
where $T$ is the tachyon field and $V(T)$ is its potential, with $V(-T)
=V(T)$ and $V(T)$ has a maximum at $T=0$, while $V(T) \to 0$ as $T \to
\pm \infty$. $\mu,\,\nu = 0,1,\ldots,p+1$ are the world-volume indices. We
have set the world-volume gauge fields and the transverse scalars to zero
for simplicity and will include them later. Also, we assume that classically
the tachyon is dependent on the time coordinate $x^0$. The equation of
motion following from (1) is,
\be
\partial_0\left[\frac{V(T) \partial_0 T}{\sqrt{1-(\partial_0 T)^2}}\right]
+ V'(T) \sqrt{1-(\partial_0 T)^2} = 0
\ee
Here `prime' denotes the derivative of the function with respect to its 
argument. Now instead of solving this equation directly if we Wick rotate
$x^0 \to i\tau$, then in terms of $\tau$ coordinate (2) can be rewritten
as 
\be
\partial_\tau\left[\frac{V(T) \partial_\tau T}{\sqrt{1+
(\partial_\tau T)^2}}\right]
- V'(T) \sqrt{1+(\partial_\tau T)^2} = 0
\ee
The solution for this equation has the form $T = f(a\tau)$, where $a$ is
a parameter which will be taken to infinity at the end. The function $f$
satisfies $f(-u)=-f(u)$, $f(\pm\infty)=\pm\infty$ and $f'(u) > 0$ for all
$u$, otherwise it is an arbitrary function. This solution is obtained by Sen
\cite{asfour} and is used to derive the codimension one BPS 
D-brane action as the
space-like tachyonic kink solution of the non-BPS D-brane action. Now we note
that if $f(a\tau)$ is a solution to eq.(3), then $f(-iax^0)$ is a solution to
eq.(2). We point out that for real $x^0$ although $f(-iax^0)$ is a solution
to eq.(2), it is unphysical as it does not satisfy the proper boundary 
condition of the time dependent tachyon. Namely, from the conservation of 
energy momentum tensor following from (1), $\partial_0 T_{00} = 0$, where
\be
T_{00} = \frac{V(T)}{\sqrt{1-(\partial_0 T)^2}}
\ee
we find $\partial_0 T \to 1$ as $x^0 \to \pm\infty$ and $\partial_0 T = 0$ at
$x^0=0$ i.e. at the top of the potential where the tachyon starts rolling.
However, we can still formally use the solution $T=f(-iax^0)$ and mention
how we can finally obtain the action in real time.

Now we consider the dynamics of the translational zero mode along $x^0$
direction and it corresponds to the fluctuation of the tachyon as,
\be
T(x^0, \xi^\alpha) = f(-ia(x^0 - X^0(\xi^\alpha)))
\ee
where $\xi^\alpha$, with $\alpha=1,\ldots,p+1$ are the world-volume 
coordinates excluding time and $X^0(\xi^\alpha)$ is a scalar living in 
$(p+1)$-dimensional Euclidean world-volume associated with the translational
zero mode of the time-like kink along $x^0$. Using (5) we find,
\be
\sqrt{-{\rm det}(\eta_{\mu\nu}+ \partial_\mu T \partial_\nu T)}
= \sqrt{1+\eta_{\mu\nu} \partial_\mu T \partial_\nu T} = \left[1+a^2f'^2(
1-\delta^{\alpha\beta}\partial_\alpha X^0 \partial_\beta X^0)\right]^{1/2}
\ee
Now substituting (6) into (1) we find for $a\to \infty$,
\bea
S &=& -\int dx^0\int d^{p+1}\xi V(f) af'\left[1-\delta^{\alpha\beta}
\partial_\alpha X^0 \partial_\beta X^0\right]^{1/2}\nn
&=& - i \int V(y) dy \int d^{p+1}\xi \sqrt{{\rm det}(\delta_{\alpha\beta}
- \partial_0 X^0 \partial_\beta X^0)}
\eea
In writing the second line in (7), we have made a change of variable
$f(-ia(x^0 - X^0(\xi^\alpha))) = y$. We notice that the integral $-i\int
V(y) dy$ is nothing but the action $S$ per unit $(p+1)$-dimensional 
Euclidean world-volume with the tachyon taking its classical value $T_{cl}
= f(-iax^0)$. However in real time the integral does not make sense as
we have mentioned before and we have to really `undo' the effect of Wick
rotation of $x^0$ coordinate by replacing $f(-iax^0) \to T_{cl}(x^0)$.
So, we have to replace,
\bea
-i\int V(y) dy &=& -\int dx^0 V(f) af'\,\,=\,\,
-\int dx^0 V(f) \sqrt{1-(-ia)^2 f'^2}\nn
&\to & -\int dx^0 V(T_{cl}) \sqrt{1-(\partial_0 T_{cl})^2}\,\,\equiv\,\, S_0
\eea
So, in real time the action would take the form,
\be
S = S_0 \int d^{p+1} \xi \sqrt{{\rm det}(\delta_{\alpha\beta} - 
\partial_\alpha X^0 \partial_\beta X^0)}
\ee
where $S_0=-\int dx^0 V(T_{cl})\sqrt{1-(\partial_0 T_{cl})^2}$. Since the
expression inside the square root in (9) is already expressed in real scalar
$X^0$, this is the DBI action of the space-like $p$ brane. Note that 
the kinetic term
of the scalar $X^0$ has a wrong sign since this is the zero mode associated 
with the time translation. 

Following Sen \cite{asfour}, it is not difficult to include world-volume 
gauge fields
and other transverse scalars into the action and we will use the same
procedure as discussed above. The action now takes the form,
\be
S = - \int d^{p+2} x V(T) \sqrt{-{\rm det} (a_{\mu\nu})}
\ee
where,
\bea
a_{\mu\nu} &=& \eta_{\mu\nu} + \partial_\mu T \partial_\nu T + \partial_\mu
x^I \partial_\nu x^I + f_{\mu\nu}\nn
{\rm with},\,\, f_{\mu\nu} &=& \partial_\mu a_\nu - \partial_\nu a_\mu
\eea
Here, $\mu,\nu = 0,1,\ldots,p+1$ are the world-volume indices of the non-BPS
D$(p+1)$-brane and $I=p+2,\ldots,9$ are the transverse space indices.
$a_\mu$ is the world-volume gauge field and $x^I$ are the scalars corresponding
to the transverse coordinates. We assume for simplicity that classically
both the gauge fields and the scalars on the world volume vanish and the 
fluctuations of various fields take the forms,
\bea
T(x^0,\xi^\alpha) &=& f(-ia(x^0-X^0(\xi^\alpha)))\nn
a_0(x^0,\xi^\alpha) &=& 0, \quad a_\alpha(x^0,\xi^{\alpha}) \,\,\,=\,\,\, 
A_\alpha(\xi^\alpha), \quad x^I(x^0,\xi^\alpha)\,\,\,=\,\,\, X^I(\xi^\alpha)
\eea
Note from above that we are assuming that the $(p+2)$-dimensional fields
$a_\mu$ and $x^I$ do not depend on time $x^0$ and the fluctuations away
from the time-like tachyonic kink are arbitrary. This makes sense for the
Wick rotated (or Euclideanized) theory and the justification can be found
in ref.\cite{asfour}. Now for this field configuration we can compute 
various components
of $a_{\mu\nu}$ as
\bea
a_{00} &=& -1-a^2f'^2, \qquad a_{0\alpha}\,\,\,=\,\,\,a_{\alpha 0}\,\,\,=\,\,\,
a^2 f'^2 \partial_\alpha X^0\nn
a_{\alpha\beta} &=& (1-a^2f'^2)\partial_\alpha X^0 \partial_\beta X^0 +
A_{\alpha\beta}\nn
{\rm where},\,\,\, A_{\alpha\beta} &=& \delta_{\alpha\beta} - \partial_\alpha
X^0 \partial_\beta X^0 + \partial_\alpha X^I \partial_\beta X^I + 
F_{\alpha\beta}
\eea
with $F_{\alpha\beta}=\partial_\alpha A_\beta - \partial_\beta A_\alpha$. Now
using (13) we find,
\be
\sqrt{-{\rm det}(a_{\mu\nu})} = a f' \sqrt{{\rm det}(A_{\alpha\beta})},
\qquad {\rm for}, \quad a \to \infty
\ee
Substituting (14) into the action (10) we get,
\bea
S &=& - \int dx^0 \int d^{p+1} \xi V(f) af' \sqrt{{\rm det}(A_{\alpha\beta})}
\nn
&=& -i\int V(y) dy \int d^{p+1} \xi \sqrt{{\rm det} (A_{\alpha\beta})}
\eea
where $A_{\alpha\beta}$ is given in eq.(13). Again in the last line we have
introduced the variable $y = f(-ia(x^0-X^0(\xi^\alpha)))$. As before we can
write the action in real time by replacing
\be
-i\int V(y) dy \to
-\int dx^0 V(T_{cl}) \sqrt{1-(\partial_0 T_{cl})^2}\equiv S_0
\ee
Where $S_0$ is the action per unit $(p+1)$-dimensional Euclidean volume
evaluated with the classical values of the fields. Therefore, we get
\be
S = S_0 \int d^{p+1}\xi \sqrt{{\rm det}(A_{\alpha\beta})}
\ee
This is the form of the DBI part of the space-like $p$-brane action. We 
note that the
first three terms in $A_{\alpha\beta}$ given in (13) i.e. 
$\delta_{\alpha\beta} - \partial_\alpha X^0 \partial_\beta X^0 + 
\partial_\alpha X^I \partial_\beta X^I$ is the pull-back of the space-time 
metric on the $(p+1)$-dimensional Euclidean world-volume of the space-like
$p$-brane
in the static gauge and so, the DBI action really has the form
\be
S = S_0 \int d^{p+1}\xi \sqrt{{\rm det}(g_{\alpha\beta} + F_{\alpha\beta})}
\ee
Taking into account the closed string background in the NSNS sector the action 
would take the form,
\be
S = S_0 \int d^{p+1}\xi e^{-\phi} \sqrt{{\rm det}(g_{\alpha\beta} + 
B_{\alpha\beta} + F_{\alpha\beta})}
\ee
where $\phi$ is the dilaton, $g_{\alpha\beta}$ and $B_{\alpha\beta}$ are 
respectively the pull-backs of the 
space-time metric and the antisymmetric tensor to the world-volume
of the space-like $p$-brane.

Apart from the DBI part the S$p$-brane (or E$p$-brane) action should 
also contain a 
Wess-Zumino term. The Wess-Zumino term of a non-BPS D$(p+1)$-brane has the 
form,
\be
S_{WZ} = \int W(T) dT\wedge c \wedge e^f
\ee
where $W(T)$ is an even function of $T$, which vanishes as $T \to \pm\infty$.
$f=f_{\mu\nu}dx^\mu \wedge dx^\nu$ and $c=\sum_{q\geq 0}c^{(p+1-2q)}$, where
$c^{(p+1-2q)}$ are the pull-backs of the RR $(p+1-2q)$-form fields on to
the world-volume. Note that here we are considering only the bosonic sector
and in the absence of the RR background the WZ term would vanish\footnote{
In fact if we include the fermionic sector $c$ also has a contribution from
the fermions and so the WZ term would be non-vanishing even if the RR 
background is zero \cite{bddep}.}. Now to evaluate (20) we first find,
\be
f = f_{\mu\nu} dx^\mu \wedge dx^\nu = 2 f_{0\alpha} dx^0 \wedge d\xi^\alpha
+ F_{\alpha\beta} d\xi^\alpha \wedge d\xi^\beta
\ee
For the fluctuation (12) the above simply reduces to $F$. Now since $dT=
-iaf'du$, where $u=x^0-X^0(\xi^\alpha)$, we can write
\be
c = \sum_{m} c^{(m)} = \sum_m \left(m c_{0\alpha_2\ldots \alpha_m}^{(m)}
\partial_{\alpha_1} X^0 + c_{\alpha_1\ldots \alpha_m}^{(m)}\right)
d\xi^{\alpha_1} \wedge \ldots \wedge d\xi^{\alpha_m}
\ee
The term in the r.h.s. of (22) is the pull-back of an $m$-form onto the
$(p+1)$-dimensional Euclidean world-volume of the space-like $p$-brane and 
can be
identified with $C^{(m)}$. Using these relations the WZ term in (20)
simplifies to
\bea
S_{WZ} &=& \int W(f) (-iaf') du \wedge C \wedge e^F\nn
&=& \int W(y) dy \int C\wedge e^F
\eea
where in writing the last expression we have introduced a new variable
$y=f(-iau)$. If we now identify $\int W(y) dy = \int V(y) dy$ in analogy with
the space-like tachyonic kink solution then
\be
S_{WZ} = i (-i\int V(y) dy) \int C \wedge e^F
\ee 
Now as argued before the integral in the paranthesis makes sense only in the
Wick rotated theory i.e. when $x^0$ is imaginary. However, if we want to
write it in real $x^0$ the integral $-i\int V(y) dy$ should be replaced by
$S_0 = - \int dx^0 V(T_{cl})\sqrt{1-(\partial_0 T_{cl})^2}$. Also if we 
include the closed string NSNS field $B$, then $F$ should be replaced by 
$F + B$, where $B_{\alpha\beta}$ should be the pull-back of the space-time 
field onto
the world-volume of space-like $p$-brane. So, the full space-like $p$-brane 
action containing the 
DBI part and the WZ part is given by
\be
S_{DBI} + S_{WZ} = S_0 \int d^{p+1}\xi e^{-\phi} \sqrt{{\rm det} (g_{\alpha
\beta} + B_{\alpha\beta} + F_{\alpha\beta})} + i S_0 \int C\wedge e^{F+B}
\ee
Thus we find that in type II theory where the RR form-fields $C$'s are 
real,
the above action is complex. So, the S$p$-brane action obtained this way has
a complex structure. Note that the action in (25) differs from the usual BPS
D-brane action by an overall factor of $i$ and therefore the equations of 
motion remain the same. This is important for their solutions to give a 
consistent background preserving the conformal symmetry of the open string
world-sheet. On the other hand, if we interpret the action (25) as that of
an E$p$-brane in type II$^\ast$ theory then the RR form fields $C$ is replaced
by $C \to C' = -iC$ \cite{ch} and the action is real. The DBI part of 
the above action
matches exactly with the S$p$-brane action obtained earlier \cite{hhw} 
using different
method. However, since in that derivation the closed string background was not
taken into account, the WZ term was absent and it was not clear whether the
action really corresponded to an S$p$-brane or an E$p$-brane action. Here, 
we observe that if we insist on the reality of the full action, then the 
action in ref.\cite{hhw} should be considered as an E$p$-brane action.

To further illustrate the possible nature of the world volume action of
the previously discussed time dependent configurations, let us consider
the following scenario. Consider the static black $p$-brane solutions of
type II supergravities. These solutions are typically singular and the
singularity is hidden behind the horizon. Due to the presence of
the horizon, one can associate a non-zero temperature with these
solutions. As a result, they completely break the original supersymmetry
of
the type II theories. It is well known \cite{lw} that, as one crosses the 
horizon of
such a solution, the role of time and space gets
interchanged. Consequently, a
static metric turns into a time dependent metric as long as we restrict
our attention inside the horizon. This time dependent metric is typically
the S-brane metric that we have been discussing so far. In the following,
we would like to consider D$p$ brane probes in such a $p$-brane
background. As we will see, the probe
brane action can be interpreted as a space-like $p$ brane action once it
crosses the 
horizon of the background geomerty. However, the action turns out to be
complex as before. To get a real action, one needs to make the
corresponding RR form of the corresponding D$p$ brane purely
imaginary. This, in turn, turns the background to a solution of a type
II$^\ast$ theory along with the probe action to the one of E$p$-brane
action. In the following, we discuss this in some detail.

The static non-extremal $p$-brane solutions of our interest are given in
\cite{dlp}. In $D$ space-time dimension, they have the form
\begin{equation}
ds^2 = e^{2A}(- e^{2f} (dx^0)^2 + dx^idx^i ) + e^{2B} (e^{-2f} dr^2 + r^2
d\Omega^2),
\label{psol}
\end{equation}
where $(x^0,x^i)$ parametrise the $p+1$-dimensional world-volume of the 
$p$-brane. The coordinates transverse to the brane are $r$ and the 
$(D-p-2)$ coordinates on the unit sphere $d\Omega$. The functional form of
$A$,
$B$ and $f$ can
be found in \cite{lmp}. For our purpose, we only need to know that they depend
solely on the radial coordinate $r$. This solution has a singularity at
$r=0$. Furthermore, $e^{2f}$ becomes zero  at finite non-zero value of
$r$. The location repersents the horizon. 
Beside the metric, there is a nontrivial dilaton $\phi$ which also is a
function of $r$. The other non-trivial field is a form field whose field
strength is given by $F = \lambda \epsilon_{D-p-2}$. Here $\epsilon_{D-p-2}$ 
is the
volume form on the unit sphere $d\Omega$. The solution (\ref{psol}) then
corresponds to a solitonic $p$-brane with magnetic
charge $\lambda$.

As discussed in \cite{lmp}, the metric in (\ref{psol}) becomes a time-dependent 
one once we consider the interior region of the $p$-brane. Inside the 
horizon, 
the function $e^{2f} \rightarrow -e^{2f}$. Consequently, as can be seen
from (\ref{psol}), the timelike coordinate $x^0$ becomes spacelike and the
radial coordinate $r$ aquires a new interpretation as the timelike coordinate. 
Renaming, therefore, the new spacelike coordinate $x^0$ as $z$ and
the timelike coordinate $r$ as $\tau$, we get the metric as
\begin{equation}
ds^2 = e^{2A}( e^{2f} dz^2 + dx^idx^i ) - e^{2B-2f} d\tau^2 + e^{2B}\tau^2
d\Omega^2,
\label{spsol}
\end{equation}
where $A, B$ and $f$ are all now functions of time $\tau$. The
functional form of the dilaton field $\phi$ remains the same as before
except that it now aquires a time dependence due to the replacement of $r$
by $\tau$. The above configuration can  be identified as a spacelike
brane configuration with $p+1$ dimensional world volume parametrised by $(z,
x^i)$. However, this brane is anisotropic on the world-volume due to the 
appearance of $e^{2f}$ only infront of $z$ coordinate. However, this
will not be of importance on what we discuss in the following.

Let us now consider a probe brane of $p+1$ dimensional world-volume in the 
background geometry given in (\ref{psol}). The action of the probe brane
will have a DBI part and the WZ part given by
\begin{equation}
S = -{\cal T}_{p+1}\int d^{p+1} \xi e^{-\phi}{\sqrt{-{\rm det}(g_{\mu \nu})}} +
{\cal T}_{p+1}\int_{{\cal{M}}_{p+1}} C_{p+1},
\label{ac}
\end{equation}
where $\mu, \nu$ run over the $p+1$ world-volume coordinates $\xi^0, \xi^1,
\ldots, \xi^{p}$ and $C_{p+1}$
is the usual RR-form associated with the brane. In the above, the RR form
is integrated over the $p+1$ dimensional world-volume. ${\cal T}_{p+1}$ is 
related to the
tension of the brane. We will now explicitly evaluate the probe action 
(\ref{ac}) in the $p$-brane background given in (\ref{psol}). We will
first consider the DBI part and later focus on the WZ part of the action.
In the static gauge
\begin{equation}
\xi^0 = x^0, ~~{\rm and}~~\xi^i = x^i,
\label{stat}
\end{equation}
the components of the induced metric $g_{\mu\nu}$ take the form
\begin{eqnarray}
g_{00} &=& -e^{2A+2f} + e^{2B-2f}\Big({\partial r\over{\partial
\xi^0}}\Big)^2
+ r^2 e^{2B} {\partial \theta_a\over{\partial \xi^0}}{\partial
\theta^a\over{\partial \xi^0}}\nonumber\\
g_{0i} &=& e^{2B -2f} {\partial r\over{\partial \xi^0}}
{\partial r\over{\partial \xi^i}} +  r^2 e^{2B} {\partial
\theta_a\over{\partial \xi^0}}{\partial\theta^a\over{\partial
\xi^i}}\nonumber\\
g_{ii} &=& e^{2A} + e^{2B-2f} \Big({\partial r\over{\partial
\xi^i}}\Big)^2 +
 r^2 e^{2B}{\partial\theta_a\over{\partial\xi^i}}
{\partial\theta^a\over{\partial\xi^i}}\nonumber\\
g_{ij} &=& e^{2B-2f}{\partial r\over{\partial \xi^i}}
{\partial r\over{\partial \xi^j}} + r^2 e^{2B}
{\partial\theta_a\over{\partial\xi^i}}{\partial\theta^a\over{\partial\xi^j}}
\label{induceone}
\end{eqnarray}
where $\theta^a$ are the coordinates on $d\Omega$. Now, we would like to
continue the probe action beyond the horizon, which occurs at the 
point where $e^{2f} =0$. As discussed earlier, this is done by
substituting 
\begin{equation}
e^{2f} \rightarrow -e^{2f}, x^0\rightarrow z~{\rm and}~r\rightarrow \tau
\label{change}
\end{equation}
in 
(\ref{psol}). However, in order to maintain the previous gauge choice
(\ref{stat}), we now need to have the following identifications of
world-volume coordinates:
\begin{equation}
\xi^0 \rightarrow \xi^{p+1},~~{\rm with}~~\xi^{p+1} = z, ~\xi^i = x^i,
\label{newstat}
\end{equation}
so that the induced metric components on the world-volume are now
\begin{eqnarray}
g_{p+1,p+1} &=& e^{2A+2f} - e^{2B-2f}({\partial \tau\over{\partial 
\xi^{p+1}}})^2
+ \tau^2 e^{2B} {\partial \theta_a\over{\partial \xi^{p+1}}}{\partial
\theta^a\over{\partial \xi^{p+1}}}\nonumber\\
g_{p+1,i} &=& -e^{2B -2f} {\partial \tau\over{\partial \xi^{p+1}}}
{\partial \tau\over{\partial \xi^i}} +  \tau^2 e^{2B} {\partial
\theta_a\over{\partial \xi^{p+1}}}{\partial\theta^a\over{\partial
\xi^i}}\nonumber\\
g_{ii} &=& e^{2A} - e^{2B-2f} ({\partial \tau\over{\partial \xi^i}})^2 +
 \tau^2 e^{2B}{\partial\theta_a\over{\partial\xi^i}}
{\partial\theta^a\over{\partial\xi^i}}\nonumber\\
g_{ij} &=& -e^{2B-2f}{\partial \tau\over{\partial \xi^i}}
{\partial \tau\over{\partial \xi^j}} + \tau^2 e^{2B}
{\partial\theta_a\over{\partial\xi^i}}{\partial\theta^a\over{\partial\xi^j}}.
\label{newind}
\end{eqnarray}
It can now easily be checked that the above components follow from the 
probe action
\begin{equation}
-i{\cal T}_{p+1} \int d^{p+1} \xi e^{-\phi}{\sqrt{{\rm det}(g_{\alpha \beta})}}
\label{dbisp}
\end{equation}
when evaluated on the background (\ref{spsol}). Here, the indices 
$\alpha, \beta$ run over the Euclidian world-volume coordinates 
$\xi^1,\ldots,
\xi^{p+1}$. This is precisely the form of space-like brane action 
as obtained before and also suggested in
\cite{hhw} if we identify $-i{\cal T}_{p+1} = S_0$, where $S_0$ is the 
action given
before in (8). Note that with this identification we encounter two puzzles.
First of all, $S_0$ as given in (8) is real, whereas $-i{\cal T}_{p+1}$
appears to be purely imaginary\footnote{We would like to point out that here
we are
assuming that ${\cal T}_{p+1}$ will remain real as we interchange $r
\rightarrow \tau$ and $x^0 \rightarrow z$, but this is not really true as
we have seen in our earlier discussion.} since ${\cal T}_{p+1}$ is related 
to the tension of the brane given in (28). Secondly, $S_0$ is the action per
unit $(p+1)$-dimensional Euclidean volume of a non-BPS brane with the tachyon
and other fields taking their classical values, but in this derivation
tachyon does not appear explicitly. The resolution of the puzzles can be
understood as follows. Note that the action (28) can also be regarded as a
non-BPS D$(p+1)$-brane action on its space-like tachyonic kink where the
tachyon as well as the other backgrounds depend only on the brane direction
$x^{p+1}=x$ (say). So, ${\cal T}_{p+1} = \int V(y) dy$ (as mentioned before)
with $y=T=f(ax)$, with $a\to \infty$ and the function $f$ is as defined 
after eq.(3). Now as the probe brane is taken inside the horizon, 
$x \rightarrow \tau$ and $x^0 \rightarrow z$, i.e. the space-like coordinate 
$x$ becomes time-like and so, $f(ax) \to f(a\tau)$, but it is no longer a 
solution to the
tachyon equation of motion. It will be a solution if we Euclideanize $\tau$
i.e. $f(-ia\tau)$ will be a solution. This is exactly the solution we used 
before. So, the factor in front of the DBI part of the action becomes 
$-i{\cal T}_{p+1} = -i \int V(y)dy$ and this when continued to the real time
gives $S_0$ (see eq.(8)).

Let us now look at
the WZ part. We note that under (\ref{change}) and
(\ref{newstat}), the WZ term in (\ref{ac}) takes the following form
\begin{equation}
{\cal T}_{p+1}\int_{{\cal M}_{p+1}} C_{p+1} \rightarrow 
{\cal T}_{p+1}\int_{{\tilde{\cal{M}}}_{p+1}} C_{p+1}.
\label{nwz}
\end{equation}
Here, ${\cal M}_{p+1}$ is the world-volume of the static brane with
coordinates $\xi^0,\xi^1,\ldots,\xi^{p}$ where as ${\tilde{\cal{M}}}_{p+1}$ 
is the world-volume of the space-like brane with coordinates $\xi^1,\ldots,
\xi^{p}, \xi^{p+1}$. Here again we identify ${\cal T}_{p+1} = i S_0$ and
consequently, the total world-volume action of the 
time dependent configuration given in (\ref{spsol}) is the sum of 
(\ref{dbisp}) and (\ref{nwz}) exactly the same form as obtained earlier
in (25). The action is clearly complex unless we
make the RR field imaginary $C_{p+1} \rightarrow -iC_{p+1}$. As discussed 
before,
under such a transformation, the solution (\ref{spsol}) really corresponds
to an E$p$ brane solution of type II$^\ast$ theory.

To conclude, we have given a simple derivation of a space-like brane 
action
including the DBI and the WZ parts starting from the non-BPS D-brane action.
We have used the time-like tachyonic kink solution of the tachyon effective
action describing the dynamics of the non-BPS D-brane by a Wick 
rotation of the
space-like tachyonic kink solution. The
fluctuations of various bosonic fields on the Euclidean world-volume around
this solution produce an effective action. In real time, this describes 
the 
low energy effective action of a space-like brane. We notice that if we start
from type II theory where the closed string background has real RR form fields
then the action is complex and it is the action of an S$p$-brane. But, 
if we start 
from type II$^{\ast}$ theory where the RR form is still real but differ from
those of type II theory by a factor of $i$ then the action is real and it
represents the action of an E$p$-brane. We pointed out that by just studying 
the
DBI part we can not conclude whether it represents the action of an S$p$-brane
or an E$p$-brane. To further support the correctness of the form of the action 
we have also given an alternative argument. 
\\
\noindent{\bf Acknowledgements:} We would like to thank Ashoke Sen for a
useful communication at an early stage of this work. We would also like
to thank Koji Hashimoto for many useful correspondences. SM would like to
thank Saha Institute of Nuclear Physics for hospitality during the course
of this work.


\end{document}